\documentstyle[11pt,aaspp4]{article}

\begin{document}

\title{On Formation and Evolution of Disk Galaxies: Cosmological
  Initial Conditions and the Gravitational Collapse}

\author{Vladimir Avila-Reese\altaffilmark{1}} \affil{Instituto de
  Astronom\'\i a-UNAM, A.P. 70-264, 04510 M\'exico D. F., M\'exico}
\author{Claudio Firmani\altaffilmark{2}} \affil{Centro de
  Instrumentos-UNAM, A.P. 70-186, 04510 M\'exico D. F., M\'exico; and
  \\ Instituto de Astronom\'\i a-UNAM, A.P. 70-264, 04510 M\'exico D.
  F., M\'exico} \and \author{Xavier Hern\'andez\altaffilmark{3}}
\affil{Instituto de Astronom\'\i a-UNAM, A.P. 70-264, 04510 M\'exico
  D. F., M\'exico}


\altaffiltext{1}{also Centro de Instrumentos-UNAM;
  avila@astroscu.unam.mx}
\altaffiltext{2}{firmani@aleph.cinstrum.unam.mx} \altaffiltext{3}{also
  Institute of Astronomy, Cambridge University, England;
  xavier@ast.cam.ac.uk}

\begin{abstract}

  We use a semianalytical approach, and the standard $\sigma _8=1$
  cold dark matter (SCDM) cosmological model to study the
  gravitational collapse and virialization, the structure, and
  the global and statistical properties of isolated dark matter (DM)
  galactic halos which emerge from primordial Gaussian fluctuations.
  Firstly, from the statistical properties of the primordial density
  fluctuation field the possible mass aggregation histories (MAHs) are
  generated. Secondly, these histories are used as the initial
  conditions of the gravitational collapse. To calculate the structure
  of the virialized systems we have generalized the secondary infall
  model to allow arbitrary MAHs and internal thermal motions. The
  average halo density profiles we obtained agree with the profile
  derived as a fitting formula to results of N-body cosmological
  simulations by Navarro et al. (1996, 1997). The comparison of the
  density profiles with the observational data is discussed, and some
  possible solutions to the disagreement found in the inner regions
  are proposed.

  The results of our approach, after considering the gravitational
  dragging of the baryon matter that forms a central disk in
  centrifugal equilibrium, show that the empirical Tully-Fisher (TF)
  relation and its scatter can be explained through the initial
  cosmological conditions, at least for the isolated systems. The
  $\sigma _8=1$ SCDM model produces galaxies with high velocities when
  compared to observations, but when the SCDM power spectrum is
  normalized to $\sigma _8=0.57$ an excellent agreement with the
  observable TF relation is found, suggesting that this relation is
  the natural extension to galactic scales of the observed galaxy
  distribution power spectrum. The theoretical TF scatter is close to
  the measured one. The slope of the TF relation is practically
  invariant witn respect to the spin parameter $\lambda $.

\end{abstract}


\section{Introduction}

A point of major interest in the study of the origin and evolution of
galaxies is the link between the physical conditions of the early
universe and the properties of present-day galaxies. This point
becomes particularly relevant if the properties of galaxies have
indeed been strongly affected by the cosmological initial conditions.
The aim of the present paper is to explore this question studying the
gravitational collapse of primordial density fluctuations, and the
characteristics of the virialized structures as a function of the
properties of the primordial density fluctuation field (hereafter
DFF). The building up and evolution of luminous disks will be treated
in a future paper.  Currently the most predictive scenario for cosmic
structure formation is based on the gravitational paradigm, and on the
inflationary cold dark matter (CDM) cosmological models. In this
scenario, to a first approximation, the problem of cosmic structure
formation can be studied by making use solely of gravitational
physics. Although N-body simulations offer the most direct method to
follow the gravitational clustering of collisionless systems, in
practice several difficulties arise. On the one hand, computational
limitations make it impossible to reach arbitrary resolutions and to
overcome problems like discreteness and collision errors (e.g.,
Kuhlman, Melott, \& Shandarin 1996; Melott et al. 1997). On the other
hand, due to the ``black box'' character of these simulations, it is
not easy to explore intermediate physical processes, and to reach a
complete statistical description of the results. All this suggests
that other alternative approaches should be developed in parallel.

To overcome the problem of the excessively steep density profiles
obtained in numerical experiments with uniform spheres collapsing in
an expanding universe, Gunn \& Gott (1972), Gott (1975), and Gunn
(1977) proposed the {\it secondary infall} process. Later, analytical
and semianalytic extensions of the secondary infall model (SIM) were
developed to study the gravitational collapse of fluctuations whose
density profiles were derived from the cosmological conditions through
the single-filter approximation. Most of them are valid only for
self-similar cases and radial infall. The first studies about the
structure of objects identified as virialized in the outcome of
cosmological numerical simulations for the Gaussian CDM models showed
a good agreement with the predictions of the SIM (see Hoffman (1988),
and the references therein). This was an amazing result because in the
numerical simulations the formation of DM halos is seen as a chaotic
process of mergers, contrary to the symmetric, centralized, and gentle
collapse implied in the SIM. Recently Zaroubi, Naim, \& Hoffman (1996)
concluded that the apparently very chaotic collapse seen in
cosmological N-body simulations seems to look ordered and gentle when
viewed in energy space. This raises the question of persistence of
memory in the initial conditions. In Voglis, Hiotelis, \& Harsoula
(1995) the gravitational collapse of spherical fluctuations with
several initial density profiles was explicitly studied by means of
N-body simulations, and a clear correlation between the profiles and
the final structures was found: the more steeply declining the initial
profiles the shallower the final structures are. Such a trend has been
confirmed by Zaroubi et al. (1996), who have pointed out that when the
initial density profiles induce a gentle collapse, order is preserved
in energy space, and the final structure reflects the initial
gravitational potential. Dark halo formation of disk galaxies may
reasonably be described by the SIM because the dynamical disk fragilty
implies that major mergers could not have played a significant role in
these cases (e.g., T\'{o}th \& Ostriker 1992).

The structure of DM halos has been explored in more detail using new
high resolution N-body cosmological simulations. The results of these
simulations suggest that the density profiles of the DM halos are
coreless and self-similar, with the mass being the simple scaling
parameter (Navarro, Frenk, \& White 1996, 1997, hereafter NWF96,
NFW97). Nevertheless, such a universal density profile is obtained as
an average fitting to the numerical results which are statistically
limited. Since the structure of DM halos could play an important role
in establishing the global properties of galaxies, a more detailed and
transparent statistical description of their characteristics is needed
in order to study galaxy formation and evolution.  Here we address the
question of {\it how do the structures of virialized objects, and
  their global and statistical properties depend on the initial
  conditions of formation defined by the primeval fluctuations}.

Tight observational structural correlations were established for disk
galaxies, as for example the luminosity-maximum rotation velocity
relation.  First, Tully \& Fisher (1977) established this relation
for luminosities in the B band, and later similar relations were found
for other passbands. A self-consistent scenario of galaxy formation
and evolution would have to explain the origin of these relations as
well as of their scatters. The question of {\it whether they are the
  result of the primordial cosmological conditions or not, is still an
  unsolved problem.} The galaxy formation and evolution approach we
propose here will allow us to investigate this question in a natural
fashion.

In Section 2, a heuristic approach aimed at connecting the
cosmological initial conditions with some global properties of
galaxies is presented. The limitations of this approach are discussed,
and overcoming them is one of the main goals of the paper. In section
3, the method we use to calculate the collapse and virialization of
spherical density fluctuations with arbitrary MAHs (initial density
profiles), and nonradial motions is described. Section 4 is devoted to
the derivation of initial conditions for the gravitational collapse
from the statistical properties of the DFF. In section 5, the results
of our calculations for the SCDM model are presented and discussed.
Subsection 5.1 specializes in the structure of DM halos; comparisons
with results obtained in N-body cosmological simulations are made, and
some possible alternatives to overcome the core problem are discussed.
Subsection 5.2 is concerned with the statistical distribution of the
different virialized configurations, their dependence on mass, and the
origin of the long wavelength Tully-Fisher (TF) relations and their
scatters. In section 6 we present the conclusions.

\section{Global galaxy properties from the primordial fluctuations: simple
  analytical comparisons}


As a first order approach, the naive top-hat spherical collapse model
(e.g., Gunn \& Gott 1972; Gott \& Rees 1975; White \& Rees 1978;
Peebles 1980) allows one to establish, for a given power spectrum of
fluctuations, a relationship between the mass $M_0$ (visible+dark) of
a r.m.s. density fluctuation and its formation (collapse) epoch $z_f$.
For a Gaussian DFF, and an Einstein-De Sitter universe:
\begin{equation}
  M_0^\beta (1+z_f)\propto const
\end{equation}
where $\beta =\frac{3+n_{eff}}6$, $n_{eff}$ is the effective index of
the power spectrum at the scale $M_0.$ For the CDM models the values
of $n_{eff}$
lie approximately between -2.0 and -2.4 for $10^{12}M_{\odot }$ and $%
10^9M_{\odot }$ systems, respectively, so that $(1+z_f)_{gal}\propto
M_0^{-\beta}$, with $\beta$ roughly between 1/6 and 1/10. Observations
show a tight relation between luminosities and line width velocities
for galaxies: the TF relation for disk galaxies ($L\propto
V_{rot}^{m_{TF}})$, and the Faber-Jackson (Faber \& Jackson 1976, FJ) for 
the elliptical ones $(L\propto V_{disp}^{m_{FJ}})$. To transform these 
empirical relations into a $M_0-V_c$ relation, where $V_c$ is the maximum
circular velocity of the dark halo before the baryon dissipative
collapse (this quantity is a measure of the gravitational potential of
the virialized structure), we assume that (1) the luminosity is
proportional to the visible mass, (2) the maximum rotation or
dispersion velocities inferred from the measured line width profiles
is proportional to $V_c$ (but see sections 4 and 5), and (3) the
visible mass is a fraction $f$ of $M_0$ and all the baryon matter is
incorporated into the visible structure (this assumption may not be
realistic especially for the low velocity systems); if the mass
distribution in galaxies resembles the primordial baryon-to-dark
ratio, $f\equiv \frac{M_{vis}}{M_0}=f_B,$ where $f_B\approx
0.05h_{0.5}^{-2}$ , and $h_{0.5}$ is the Hubble constant in units of
50 kms$^{-1}$Mpc$^{-1}$. Then, we obtain,

\begin{equation}
  M_0=\frac{\Upsilon Ck^m}{f_B}V_c^m
\end{equation}
where $C$ and m are the coefficient and the slope of the empirical TF or FJ relations, $\Upsilon $ is the mass-to-luminosity ratio, and $k$ is the factor 
by which the maximum circular velocity increases after the dissipative
collapse. Now, assuming that the radius of the virialized object is
half the radius attained at the maximum expansion, the top-hat model
gives:
\begin{equation}
M_0^{\frac 23-\frac 2m}(1+z_f)=\left( 0.013h_{0.5}^{1/3}k\left( \frac{%
  C\Upsilon }{f_B}\right) ^{1/m}\right) ^{-2}
\end{equation}
In the case of the infrared (H band for example) TF relations the
slope $m$ seems to be much smaller than the canonical value of $\sim
$4 found in early studies with aperture photometry. A value in the
range of 3.0-3.5 is more accepted today (e.g. Pierce \& Tully 1988;
Gavazzi 1993; Bernstein et al.  1994; Gavazzi, Pierini, \& Boselli
1996). Thus, according to eq. (3), $(1+z_f)_{gal}\propto M_0^{-\beta
  }$ with $\beta $ between $\sim $0 and $\sim $1/10. Therefore, for
disk galaxies the cosmological initial conditions either play a
marginal role in establishing the TF relation, and/or the very
simplified top-hat model is not able to describe the process of
formation of these galaxies. For elliptical galaxies the slope of
the FJ relation is roughly independent from the passband, and
oscillates between 3.6 and 4.0 for the B to the H bands, respectively
(de Vaucouleurs \& Olson 1982). Hence, the slope of the Faber-Jackson
relation is in agreement with the one derived from the power spectrum
according to the top-hat model. It is commonly thought that the
formation of elliptical galaxies indeed could have occurred through a
violent collapse because they are typically more concentrated objects than
disk galaxies, and poor in young stellar populations.

Another important global property of disk galaxies is the
luminosity-radius relation, $L\propto R_L^n.$ As Burstein and Sarazin
(1983) have shown, this relation and the TF relation are connected
through the protogalactic conditions. Under certain assumptions they
found that $M_0^{1/n-1/3}(1+z_f)\propto const.$ Comparing this
expression with eqs. (1) and (3) one obtains $\beta =\frac 23-\frac
{2}{m_{TF}}=\frac 1n-\frac 13$ for disk galaxies. Gavazzi et al. (1996) have
found $n\approx $2.5 for observations in the H band, which is in
excellent agreement with the observational estimates for $m_{TF}$, but
again in conflict with the values predicted from the CDM power spectra
in the frame of the top-hat model.

The most serious inconsistency of the top-hat model is related to the
fact that the formation of cosmic structures implies a {\it continuous
  process of mass aggregation}. Furthermore, a given present-day mass
$M_0$ is not characterized by a unique aggregation history, but {\it a
  wide range of possibilities exists according to the statistical
  properties of the DFF} (section 4). If, as a first order
approximation, the MAHs are characterized by one parameter that could
be, for example, the redshift when half of the mass $M_0$ is reached,
$z_f(M_0/2),$ then relation (1) might still be meaningful, and the
model presented above could be useful as a heuristic approach able
to connect initial cosmological conditions with global galaxy
properties. In what follows we shall resort to a more detailed
analysis of galaxy formation with the aim to study the questions
stated here, as well as the structure of virialized objects, and to
explore to what extent the properties of galaxies are connected with
the cosmological conditions.

\section{The virialization of dark halos}


\subsection{The analytical approximation}

The dynamics of collapsing spherical shells in an expanding universe
can be calculated in a first approximation through the spherical
collapse model (e.g., Peebles 1980). The dynamics of these shells
depend only on the mass contained within them and on the cosmological
model. In the case of an Einstein-de Sitter universe, used throughout
this paper, the collapse is described by the solution of a closed
universe embedded in a flat universe.

Once the initial mass excess (over the average density of the
universe, linearly extrapolated to the present epoch), $\Delta _o(x),$
within a sphere of commoving radius $x$ is given, the solution of the
spherical collapse model allows us to calculate the imaginary density
profile constructed by phasing out the maximum expansion times of all
the shells to a fixed time.  For simplicity consider the case of a
self-similar profile, $\Delta _o(x)=\Delta _c(\frac x{x_c})^{-\alpha} $,
where $\alpha \equiv -\frac{d\ln \Delta _o(x)}{d\ln x}=const$. After
maximum expansion a mass shell falls to the center, and in a time of
the order of the dynamical time virializes. If, as a preliminary
approximation, we consider that the virialized (time averaged) radius
is proportional to the maximum expansion radius, $r_v=Fr_m$ with
$F=const$ for a given $\alpha $, then the shape of the imaginary
density profile at the maximum expansion is conserved after the
virialization. Thus, the density profile of the DM halo will be given
by:
\begin{equation}
\rho _v(r_v)=6\times 10^5\frac{\overline{\rho }_0}{1+\alpha }\left[ \frac{%
  \left( 1.1\Delta _c\right) ^{2-\alpha }h_{0.5}^{2\alpha
      }}{F_{0.5}^3}\left( \frac{M_c}{10^9M_{\odot }}\right) ^\alpha
\right] ^{\frac 1{1+\alpha }}\left( \frac{r_v}{kpc}\right)
^{-\frac{3\alpha }{1+\alpha }}
\end{equation}
where $F_{0.5}$ was normalized to 0.5, $M_c=\frac{4\pi }3\rho
_0x_c^3,$and $\overline{\rho }_0$ is the present average density of
the universe. The case $\alpha =2$ leads to an isothermal structure,
while for $\alpha \gg 1$ a density profile decaying as $\sim r^{-3}$
arises.

Since in our approach to galaxy formation we try to reconstruct a
present-day galaxy as a function of its collapse regime, instead of
the fluctuation profiles it is more appropriate to give as initial
conditions the MAHs parametrized in such a way that a total mass $M_o$
is attained by the present epoch $t_o$:
\begin{equation}
  M(t)=M_0\left( \frac t{t_o}\right) ^\gamma
\end{equation}
In eq.(5) $t$ is the collapse time conventionally defined as twice the
maximum expansion time. Using the spherical collapse model we have for
the self-similar cumulative density profile that $M(t)=\frac{4\pi
  }3\rho _ox_c^3\left( \frac{\Delta _c}{1.69}\right) ^{\frac{3\gamma
    }2}\left( \frac t{t_o}\right) ^\gamma $ with $\gamma =2/\alpha $,
and on assuming a constant collapse factor $F$ the density profile of
the virialized object (eq. (4)) in terms of the parameters of the MAH
will be:
\begin{equation}
  \rho (r_v)\approx 6\times 10^5\overline{\rho }_o\left(
  \frac{58^{2(1-\gamma )}}{F_{0.5}^{3\gamma }h_{0.5}^4}M_{10}^2\right)
  ^{\frac 1{2+\gamma }}\left( \frac \gamma {2+\gamma }\right) \left(
  \frac{r_v}{kpc}\right) ^{-\frac 6{2+\gamma }}
\end{equation}
where $M_{10}$ is $M_0$ in units of $10^{10}M_{\odot }.$ Given the
density profile by a simple integration it is easy to calculate the
circular velocity profile:
\begin{equation}
V(r_v)\approx 28\left( \frac{58^{(1-\gamma )}h_{0.5}^\gamma }{F_{0.5}^{\frac{%
      3\gamma }2}}M_{10}\right) ^{\frac 1{2+\gamma }}\left(
\frac{r_v}{Kpc}\right) ^{\frac{\gamma -1}{2+\gamma }}kms^{-1}
\end{equation}
Immediately one sees that flat circular velocity profiles are obtained
for $\gamma \approx 1,$ the high accretion regime case, while for the
low accretion regime, $\gamma \approx 0$, the velocity profiles are
almost Keplerian. In this case one has a very concentrated object in
the central regions (the circular velocities are high toward the
center), while the most external regions are loosely bound. Eqs. (6)
and (7) are useful as a first approximation to the more complex cases
of no self-similarity and non constant collapse factors.

\subsection{A higher order approximation for the virialized structure
  calculation}

The main difficulty in calculating the virialization of spherical
fluctuations in an expanding universe is introduced by the shell
crossing which occurs after the maximum expansion phase. The problem
is much simplified assuming that the dynamical time of a given inner
shell is smaller than the outer shell aggregation time. Thus, the
inner shell admits an adiabatic invariant. Gunn (1977) was the first
who applied the adiabatic invariant formalism to calculate the effect
of secondary infall on a virialized system with a power-law density
profile. In Zaroubi \& Hoffman (1993) the exact solutions for the
asymptotic limit in the self-similar case were presented. We expand
the approach of Zaroubi \& Hoffman in order to study the more general
case of non- self-similar collapse and non-radial motion.  Ryden \&
Gunn (1987) treated this problem using another approach.

Under the assumption of spherical symmetry the fluctuation from which
a galaxy emerges can be described as a sequence of concentric shells.
During the early cosmological expansion each shell contains a mass
$m_o$, and it
will reach a first maximum expansion at radius $r_o$; the function $%
m_o\left( r_o\right) $ is given by the initial conditions. We assume
that, given a shell, the internal ones have reached the first maximum
expansion radius earlier than this, which implies a radially declining
density profile for the fluctuation. After its first maximum
expansion, each shell will fall toward the center due to the
gravitational field of the internal mass and will evolve through a
radial oscillatory movement. Each oscillation defines a maximum
expansion radius $r_a$ (apapsis), which retains the information of
$r_o$. The function $r_a\left( r_o\right) $ is crucial in our problem,
and according to the initial conditions, it is an increasing monotonic
function.  The gravitational field at a radius $r$ is given by the
shells with apapsis $r_a\leq r$ that permanently oscillate inside the
radius $r$ and whose mass is $m_P\left( r\right) $, as well as by the
shells with apapsis $r_a\geq r$ that only momentarily fall inside
the radius $r$ and whose mass contribution is $m_M\left( r\right) $.
If we assign to each shell a function $P\left( r,\xi \right) $, where
$\xi $ represents the apapsis of the shell, and $P$ the probability to
find the shell inside the radius $r$, then
\begin{equation}
  m_M\left( r\right) =\int_r^{R_o}P\left( r,\xi \right)
  \frac{dm_P}{d\xi }d\xi
\end{equation}
where $R_o$ is the apapsis of the last shell, {\it i.e.} $r<R_o$
always. The total mass inside a radius $r$ is
\begin{equation}
  m_T\left( r\right) =m_P\left( r\right) +m_M\left( r\right)
\end{equation}
The function $P\left( r,\xi \right) $ may be written as
\begin{equation}
  P\left( r,\xi \right) =\frac{\int_0^r\frac{d\eta }{v\left( \eta ,\xi
    \right) }}{\int_0^\xi \frac{d\eta }{v\left( \eta ,\xi \right) }}
\end{equation}
where $v\left( r,\xi \right) $ is the radial velocity of a shell with
apapsis $\xi $ when its radius is $r$. For a given shell the expression
of $v\left( r\right) $ is
\begin{equation}
v^2\left( r\right) =2\left[ E-G\int_0^r\frac{m_T\left( \eta \right) }{\eta ^2%
  }d\eta -\frac{j^2}{2r^2}\right]
\end{equation}
where $E$ is the total energy of the shell, and $j$ is the typical
angular momentum per unit mass of a shell mass element due to the
thermal motion which will be taken constant in time. The condition
$v=0$ defines the apapsis $r_a=r_{\max }$ and a value $r_{\min }$
which determines the maximum penetration of the shell towards the
center. We express $j$ through the ratio $e_o\equiv \left(
\frac{r_{\min }}{r_{\max }}\right) _o$ and, although it is defined for
the first $r_{\max }$ and $r_{\min }$ of every shell, through $e_o$ we
are parametrically taking into account the thermal energy that could
be produced by the mergers of substructures and tidal forces at all
times. The sequential aggregation of new shells, combined with their
motion toward the center, introduces new contributions to the
gravitational field which acts on the underlying shells. This
non-conservative spherical gravitational field changes $E$, and
consequently $r_a$ and $e$ of each shell. The contraction of $r_a$
leads to an asymptotic value of $r_a$ which is identified as the
current virialized radius $r_v.$ The change of $E$ may be estimated by
assuming an adiabatic invariant for the radial motion. The problem is
completely determined when $m_o\left( r_o\right) $ (related to the
MAH) and $j_o\left( r_o\right) $ (related to the parameter $e_o)$ are
assigned. A simple iterative numerical method allows us to calculate the
solution.

Once the initial conditions are given, through the method described
above we are able to calculate the gravitational collapse and obtain
the virialized structure of a DM halo. As thermal energy tends to zero
(almost pure radial orbits), the virialized structure tends to present
flat rotation curves down to the center, irrespective of the MAH. An
increment in the value of the thermal energy (increasing the parameter
$e_o$ for each shell) produces shallower cores in the halo. With the
aim to test the numerical method the structures obtained in the
self-similar case were compared with the analytical approximation
given by eq.(7). An excellent agreement was found (Avila-Reese 1998).

\subsection{Gravitational influence of the dissipative collapse}

Baryonic matter is able to dissipate energy, therefore virialization
is only an intermediate phase in the collapse of baryonic gas. This
collapse can be halted only by centrifugal forces and/or when gas is
transformed into stars.  Here we assume that the gas fraction of a
shell that has attained its maximum expansion, is incorporated into a
disk in centrifugal equilibrium within a dynamical time (for galaxies
the gas cooling time is typically smaller than the dynamical time
(Silk 1977; Rees \& Ostriker 1977; Silk 1984)). The collapsing baryon
mass drags DM by gravitational interaction, making the DM halo more
concentrated. We estimate this effect by allowing the baryon mass
fraction, initially distributed as DM, to form a disk in centrifugal
equilibrium, and by calculating its gravitational influence on the DM
under the assumption of adiabatic compression of the halo (see e.g.,
Flores et al. 1993). On what concerns the ordered angular momentum it
is assumed that the shells are in solid body rotation, and that every
mass element maintains its angular momentum during the radial
contraction. The angular momentum acquirement by protogalaxies is
commonly explained by the tidal torques induced by neighboring density
fluctuations on a given fluctuation that is still in its linear
gravitational collapse regime (e.g., Peebles 1969; Thuan \& Gott 1977;
Fall \& Efstathiou 1980; White 1984; Catelan \& Theuns 1996). Using
the Zel'dovich approximation it is possible to estimate how the
specific angular momentum of a shell that contains a mass $M$ depends
on this mass and on the time when the shell separates from the overall
expansion, $t_m,$ or equivalently, on the maximum expansion radius,
$r_m:$
\begin{equation}
  j\propto t_m^{1/3}M^{2/3}\propto (GMr_m)^{1/2}
\end{equation}
(e.g., White 1994). The angular momentum of structures identified as
virialized in the analytical studies and the outcomes of N-body
simulations
is commonly expressed through the spin parameter $\lambda \equiv \frac{%
  J\left| E\right| ^{1/2}}{GM_o^{5/2}},$ where $J$ and $E$ are the
total angular momentum and energy of the system. Most of these works
find an average value for $\lambda $ around 0.05 (e.g. Catelan \& Theuns 
1996, and references therein). We fix the factor
needed in relation (12) in such a way that the $\lambda $ of the
evolving dark halo is equal to 0.05.

\section{Initial conditions for galaxy formation from the primordial density
  fluctuation field}


The two initial conditions required for calculating the collisionless
gravitational collapse through the method described in subsection 3.2
are the MAH and the thermal energy content. The last one depends on
the degree of substructure of the collapsing region and on the
external tidal field.  The limitations of the method (spherical
symmetry and a not direct treatment of external forces) do not allow
to explicitly introduce these effects.  However, the thermal energy
content can be considered in the problem as a free parameter that, as
was discussed in subsection 3.2, is characterized by the orbital axis
ratio $e_o$.

In what regards the MAHs, their determination from the primeval DFF is
much more direct. In the first atempts to establish a direct
connection between the properties of virialized objects and the
cosmological initial conditions (Peebles 1984; Hoffman \& Shaham 1985;
Ryden \&\ Gunn 1987; Ryden 1988), the mean density profiles around
local maxima of Gaussian fields calculated with the single-filter peak
formalism (Peebles 1984; Bardeen et al. 1986) were used.
Notwithstanding the good qualitative results obtained in these works,
the single-filter peaks statistics is not adequate for calculating the
initial density profiles from which galaxies will emerge, because of
the high degree of correlation between the substructures on small
scales with the smoothed fluctuation (Bardeen et al.  1986; Bond \&
Myers 1996).

The Press-Schechter or excursion set approach (e.g., Press \&
Schechter 1974; Peacock \& Heavens 1990; Bower 1991; Bond et al. 1991;
Jedamzick 1995) offers a more satisfactory method for determining the
MAHs of dark halos (but see also Manrique \&\ Salvador-Sol\`{e} 1995,
1996). The key concept is that of the conditional probability that
given a collapsed region of mass $M_1$ at some time $t_1$, an object
of mass $M_2$ (spatially chosen at random from within the volume
containing the mass $M_1$) had collapsed at some earlier time $t_2.$
This probability was independently deduced by Bower (1991) and Bond et
al. (1991) for Gaussian fields. In this case the initial fluctuation
power spectrum is the only statistical information needed to
completely define the fluctuation field. The regions which have
collapsed at time $t$ are identified as those regions whose linear
cumulative density contrast $\Delta _o(M,t)$ has reached a critical
threshold $\delta _c.$ The spherical collapse model allows us to
exactly calculate this threshold for any cosmology. Assuming
$t_c=2t_m$, for an Einstein-de Sitter universe one obtains $\delta
_c\approx $1.69. A good agreement between the analytical
Press-Schechter mass distribution -where $\delta _c$ plays the role of
a parameter- and the results of N-body simulations has been found when
a value smaller than 1.69 is used. We shall use the value suggested by
Carlberg \& Couchman (1989), $\delta _c$=1.44. Now, fixing an initial
mass $M_0$ and its cumulative density contrast at $t_o$, recurrent
trials with the mass distribution given by the conditional probability
are applied (Lacey \& Cole 1993, Eisenstein \& Loeb 1996) to construct
several MAHs which will serve as initial conditions for the
gravitational collapse and virialization of dark halos. An example of
what these MAHs look like can be found in Figure 1 of Avila-Reese \&
Firmani (1997).

Since our interest is to determine the initial conditions for {\it
  disk} galaxy formation, some constraints have to be applied to the
MAHs. Disk galaxies tend to be isolated systems. That is why,
according to the definition of isolated objects given by Jedamzick
(1995), we consider only systems whose present-day mass virialized
from a region overdense by $\delta _c$ which is not yet included in a
larger region overdense by $\delta _c.$ On the other hand, in calculating
the MAHs one must follow the history of the main progenitor, 
which is attained by identifying the most massive subunity of the 
distribution at any time. This is realized as follows: for 
each temporal step at which a random mass $M_2$ at time $t_2$ was found 
given the mass $M_1$ at time $t_1$ through the Monte Carlo trial, we ask 
if the complementary mass is larger than $M_2;$ if this is the case then, 
using mass conservation, we find the cumulative density contrast of the
complementary region and again carry out a trial for obtaining $%
M_2^{^{\prime }},$ the mass contained in the complementary region and
collapsing at $t_2.$ This procedure is repeated until the current
complementary mass is smaller than the calculated collapsed mass.
Among the collapsed regions with masses $M_2,$ $M_2^{^{\prime }},...$
the most massive is chosen as the corresponding to $t_2$ and it is
used as the conditional for the next time step.

\begin{figure}
\plotone{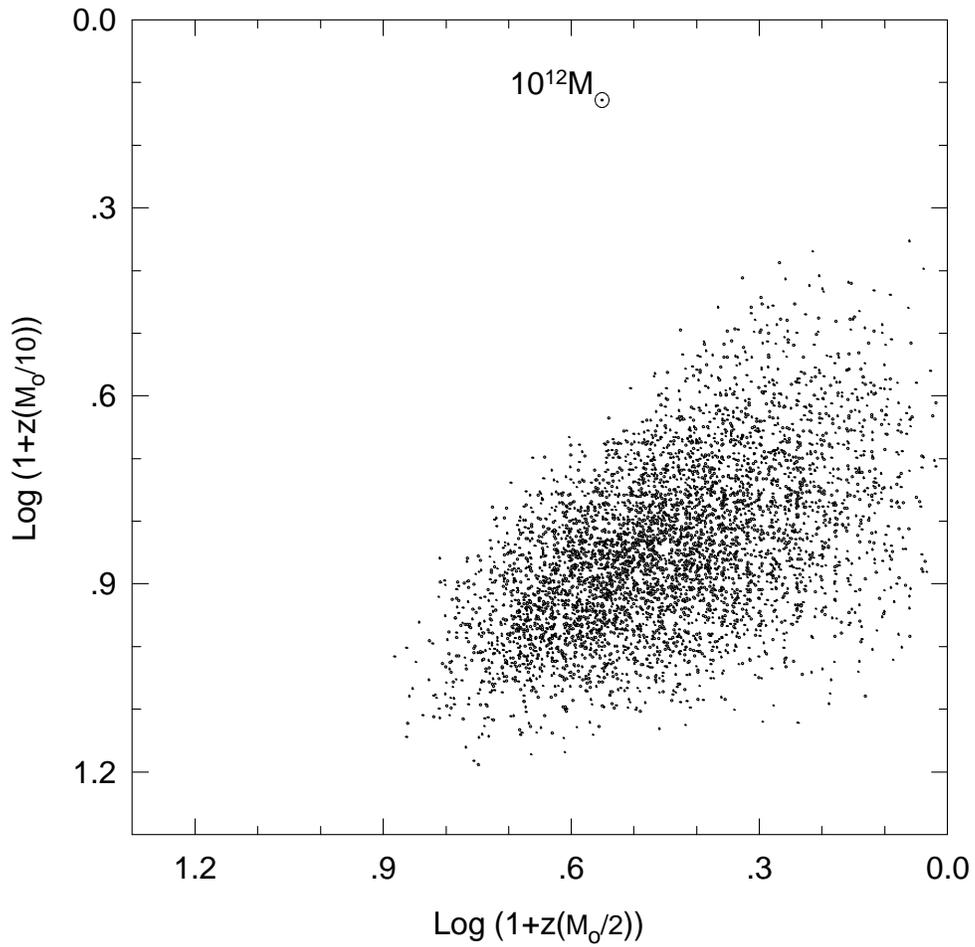}
\caption[]{Statistical distribution of the MAHs for a Gaussian SCDM
  model in the $z_1(M_0/2)-z_2(M_0/10)$ plane. $z_1$ and $z_2$ are the
  redshifts at which the mass trajectories attain one half and one
  tenth of the present-day mass $M_0=10^{12}M_{\odot }$, respectively.
  \label{fig1}}
\end{figure}

\placefigure{fig1}

The random MAHs obtained are particular realizations of a
multidimensional parent distribution defined by the statistical
properties of the DFF. To reach a statistical description of the
parent distribution in strict terms we should work in a n-dimensional
space (n is the number of temporal steps), where every mass
aggregation trajectory would have to be represented by a point.
Nevertheless, a reasonable statistical description of the MAHs can be
still given by two parameters, for example the redshifts $z_1$ and
$z_2$ at which they attain half and one tenth of the present-day mass,
respectively (a more detailed statistical analysis of this question is
treated in Avila-Reese 1998). In the plane $z_1(M_0/2)-z_2(M_0/10)$ a
point will correspond to a random MAH (see Figure 1). The random MAHs
which fall into a given small region of this plane have rather similar
morphologies along all their points, showing that two points are
enough to characterize them.

\section{Results and discussion}


We now apply the method for calculating the virialization of spherical
fluctuations in an expanding universe presented in section 3 to the
initial conditions inferred from the statistical properties of the
primeval DFF in the way described in section 4. Since our aim is to
study the general behavior of the hierarchical clustering scenario in
what concerns galaxy formation, as well as the structure and scaling
laws of galactic halos, whose generalities do not depend on the
assumed cosmology, results for a representative model are sufficient.
Here these results are presented for the SCDM model with the power
spectrum normalized to $\sigma _8=1$, which sets our work on a common
ground with other studies, and facilitates comparisons of the
approach.

\subsection{The structure of dark matter halos}

For a fixed present-day mass the initial conditions (the MAHs) will
give rise to a family of DM halo structures. We choose as the
representative cases the MAH obtained as the average of all the mass
trajectories, and two trajectories (upper and lower) corresponding
roughly to the two extreme deciles of the distribution of trajectories
in the $z_1(M_0/2)-z_2(M_0/10)$ plane. This means that roughly 80\% of
all galactic halos will fall in the range defined by the chosen upper
and lower cases. The median case is typically very similar to the
average one.

\begin{figure}
\plotone{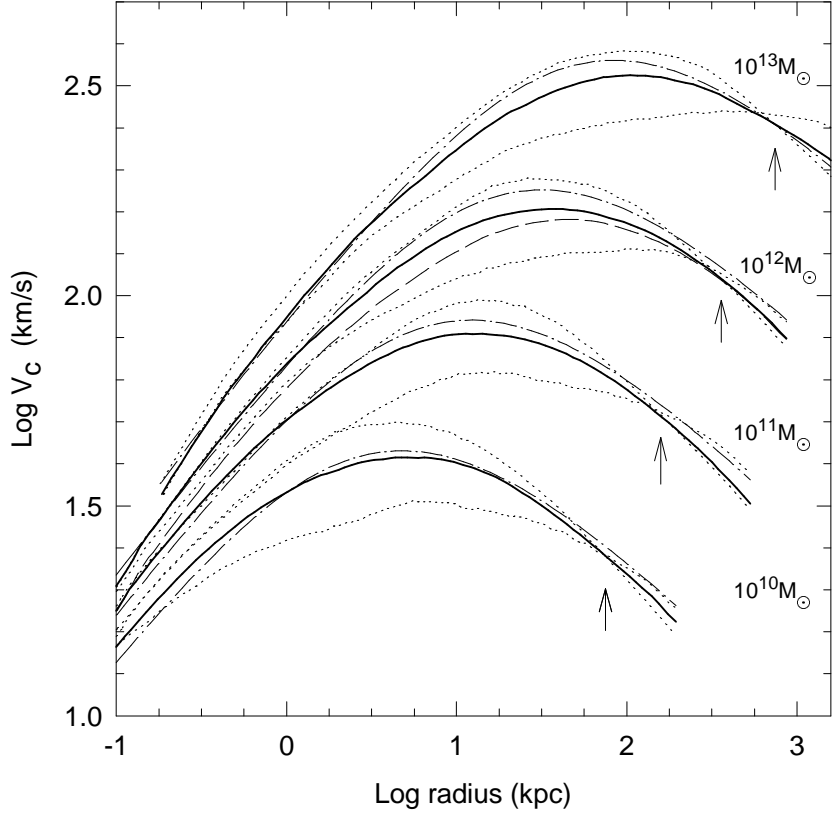}
\caption[fig2.eps]{Circular velocity profiles of $10^{10}M_{\odot
    },$ $10^{11}M_{\odot },$ and $10^{12}M_{\odot }$ DM halos where
  the average (solid lines) and low and high (upper and lower dotted
  lines, respectively) MAHs were used as the initial conditions (see
  text). The dot-dashed lines are the analytical fitting to the
  outcomes of N-body cosmological simulations given in NFW96,$\ 
  $NFW97. The arrows indicate the virialized radius $r_h$ that at the
  present epoch encompasses the given mass for the average cases; for
  the low and high aggregation cases these radii are almost
the same. The dashed line corresponds to the average MAH of a $%
10^{12}M_{\odot }$ halo but calculated with $\delta _c=1.69.$
\label{fig2}}
\end{figure}

\placefigure{fig2}

Figure 2 presents the logarithmic rotation curves of DM halos of $%
10^{10}M_{\odot },$ $10^{11}M_{\odot },$ $10^{12}M_{\odot },$ and
$10^{13}M_{\odot }$ for the three chosen MAHs. The asymptotic
structures of the halos are in agreement with the analytical solution
for the self-similar case (subsection 3.1): halos formed through an
initial fast collapse have a more concentrated structure than those
which undergo a more extended collapse. To compare the DM halo
structures obtained through our approach with the outcomes of
cosmological N-body simulations, in Figure 2 we also plot the
``universal'' velocity profile proposed in NFW96 and NFW97
(appropriately calibrated to the SCDM $\sigma _8$=1 model) (dot-dashed
lines). The circular velocity profiles inferred from these N-body
simulations not only fall in the statistical range calculated here,
but are in rough agreement with the profiles corresponding to the
average initial conditions. It is encouraging that through two very
different methods the result tends to be almost identical. This
implies ({\it i}) that the complex nonlinear processes related to the
non-sphericity of real fluctuations and the influence of external
tidal forces play only a second-order role in establishing the
structure of DM halos, and ({\it ii}) that the adiabatic invariant
assumption on which the SIM is based is a good approximation to the
problem of the cosmological gravitational collapse. As a matter of
fact processes related to the mergers and tidal forces which could
produce tangential perturbations (thermal energy) in the collapsing
matter, were implicitly considered in our models through the free
parameter $e_o$; it appears that the detailed description of these
processes is largely erased by the virialization process, remaining
only through the value of $e_o$. The value $e_o=$0.3 was fixed for all
the shells.

\begin{figure}
\plotone{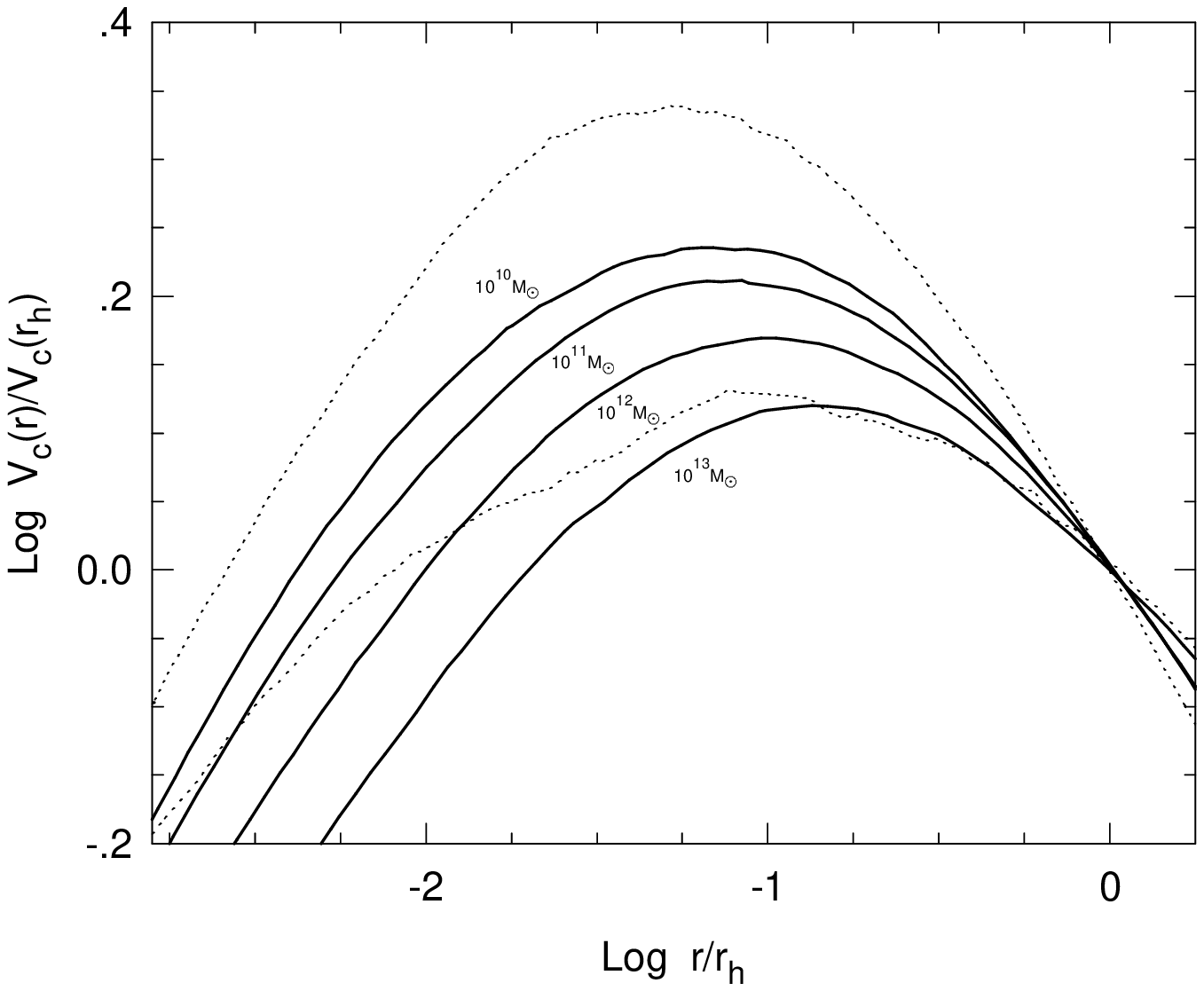}
\caption[fig3.eps]{Scaled circular velocity profiles of
  $10^{10}M_{\odot },$ $10^{11}M_{\odot }$, $10^{12}M_{\odot },$ and
  $10^{13}M_{\odot }$ DM halos corresponding to the average MAHs. The
  dotted lines are the velocity profiles of $10^{10}M_{\odot }$ halos
  for the low (upper), and high (lower) MAHs. The radius was scaled to
  the virialized radius $r_h$ that at the present epoch encompasses
  the given mass, and the velocity at this radius.
\label{fig3}}
\end{figure}

\placefigure{fig3}

In Figure 3 we present the scaled rotation curves for the case of the
most probable trajectories for four halo masses. For the $10^{10}$
$M_{\odot }$ halo the upper and lower deviations are also plotted
(dotted lines). We scale radii by the virialized radius $r_h$ at the
present epoch which encompasses $M_0$, and circular velocities by
their values at this radius.  One observes a tendency of low mass
systems to have more concentrated circular velocity profiles than the
high mass ones, although the scatter for a given mass due to the
statistical dispersion of initial conditions, is as significant as the
differences due to the mass. Therefore, the
density profiles of DM halos depend at least on two factors, namely {\it %
  (i)\ the mass and (ii)\ the scatter in the initial conditions which
  is in turn determined by the statistical properties of the
  primordial }DFF. If the formation histories and structural
properties of galactic halos are important in establishing the
properties which define the Hubble sequence then these two factors
should be considered when studying the origin of this sequence.

The fact that for the less massive halos the aggregation of matter is
typically faster at early epochs than for the more massive halos is
ultimately determined by the primordial fluctuation power spectrum. As
a rough approximation, the epoch when the protogalaxy has attained
half the present-day mass $M_0$ can conventionally be defined as its
formation epoch $z_f$ (Lacey \& Cole 1993). Then, through the
excursion set formalism the distribution of the so-defined formation
redshifts $z_f$ can be found. Using the average or median values of
this distribution a direct relation between mass and the epoch of
formation is derived (formula 2.32 in Lacey \& Cole 1993). For
instance, taking $n_{eff}$=-2, $\sigma _8$=1, the median value for
$\omega _f$ (see Lacey \& Cole), and the two masses $10^{10}M_{\odot
  }$ and $10^{13}M_{\odot },$ in the case of the top-hat model
$1+z_f\propto M_0^{-1/6},$ while according to the excursion set
formula one gets approximately $1+z_f\propto M_0^{-1/11}.$ Our
simulations for a power-law spectrum with $n_{eff}$=-2 and using the
most probable MAHs give the same dependence. For the SCDM $\sigma
_8$=1 power spectrum, such a dependence turns out even smaller,
roughly $1+z_f\propto M_0^{-1/15}.$ Now in the light of the
hierarchical clustering scenario, it is worth emphasizing that the
observed slopes of the TF and luminosity-radius relations (see section
2) are in excellent agreement with the slope $\beta \approx $1/15 of
this nominal mass-formation redshift relation.

The central structure of DM halos is affected by the gravitational
dragging of the baryon matter that dissipates and collapses within
them. Using the method described in subsection 3.3 we have calculated
this effect and obtained the final rotation curves. The values of 0.05
were used for the parameters $\lambda $ and $f_B$ (assuming $f=f_B),$
respectively, although it is important to bear in mind that these
parameters can span a range of values, as is the case of the
dispersion in $\lambda $ found in the numerical and analytical
calculations. This will introduce an extra scatter in the results of
the final (after baryon collapse) structures.  The main relevance of
such a dispersion is related to the understanding of the disk Hubble
sequence and the nature of the LSB galaxies (e.g.,
Dalcanton, Spergel, \& Summers 1996). In the $10^{10}M_{\odot }$ -$%
10^{12}M_{\odot }$ range the maximum circular velocities after the
dissipative collapse grow approximately by factors of 1.30 and 1.43
for the low and high MAH, respectively, and 1.34 for the average MAHs.
The total rotation curves for all the cases are nearly flat or
slightly decreasing.  However, the rotation curves typically rise very
steeply. This point seems to be related to a possible inconsistency of
the models: the cuspy inner structure of DM halos. Indeed, when
looking at the rotation curve decompositions, the modeled structures
show a strong contribution to the gravitational potential of DM down
to very small radii of the galaxy. The standard rotation curve
decompositions applied to normal galaxies are commonly well fitted
with a dominant disk gravitational contribution over that of the
dark halo at the central regions (e.g. van Albada et al.
1985). In Figure 4 the rotation curve decomposition is shown for a $%
10^{11}M_{\odot }$ galaxy where the most probable MAH was used. If the
original dark halo velocity profile (``halo before'' in the Figure)
would rise more slowly, then the baryonic matter contribution to the
gravitational potential would dominate at the center after its
collapse, maintaining at the same time the rotation curve nearly flat.
As we shall see, more direct observations of the internal structure of
galactic halos seem to confirm the acuteness of this problem.

\begin{figure}
\plotone{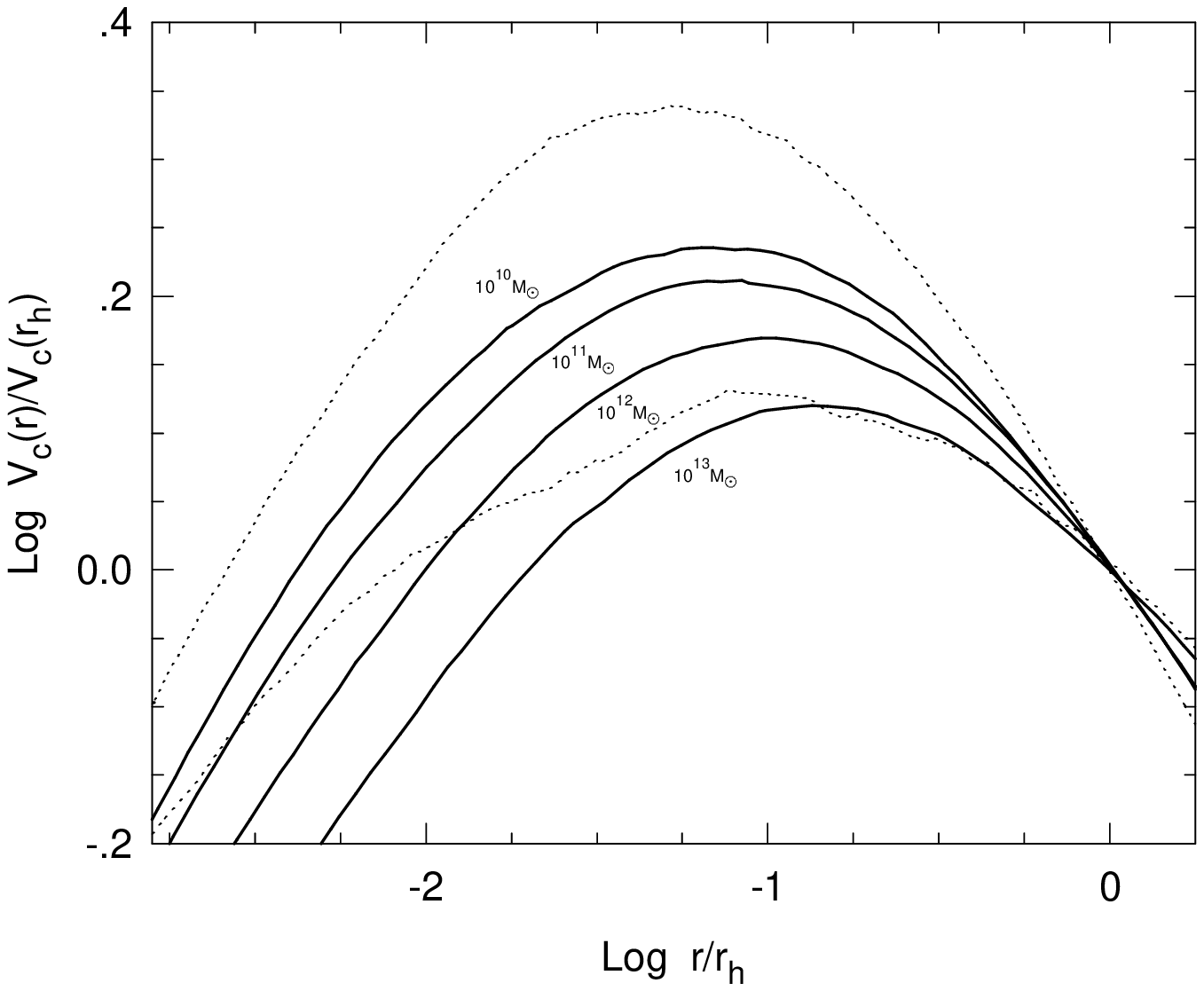}
\caption[fig4.eps]{Rotation curve decomposition of a
  $10^{11}M_{\odot }$ galaxy where the average MAH was used. The solid
  line is the total rotation curve, while the dot-dashed and dashed
  lines correspond to the dark halo and disk components, respectively.
  The two-dot-dashed line is the circular velocity profile before the
  gravitational dragging produced by the baryon matter. \label{fig4}}
\end{figure}

\placefigure{fig4}

\subsubsection{The inner density profile of dark matter halos}

Since dwarf galaxies probably are systems strongly dominated by DM,
their rotation curves could be providing direct information about the
internal structure of the ``virgin'' DM halos. Flores \& Primack
(1994), Moore (1994), and Burkert (1995) have shown that the inner
structures inferred from the rotation curves of dwarf spiral galaxies
are characterized by a large core where the density profile is nearly
flat. Low surface brightness galaxies also seem to provide direct
probes of the internal structure of their dark halos, and signatures
of large cores in them were found (de Blok 1997; de Blok \& McGaugh
1997). Flores \& Primack (1994) have also argued that gravitational
lensing analyses on cluster scales exclude the possibility of coreless
DM structures. Thus, the observational data about the internal
dynamics of those systems whose DM halos have not been much deformed
by the dissipative collapse of baryon matter tend to confirm that
nature produces DM halos with large cores. As has been shown above,
our results for a Gaussian SCDM model predict DM halos with density
profiles slightly shallower than $\rho \propto r^{-1}$ for the central
regions. From the outcomes of high resolution cosmological N-body
simulations carried out in recent years, similar inner profiles
(Dubinsky \& Carlberg 1991; Katz 1991; Warren et al. 1992; Crone,
Evrard, \& Richstone 1994; NFW96,$\ $NFW97; Cole \& Lacey 1996,
Kravtsov et. al. 1997a), or even steeper (Fukushige \& Makino 1996,
Moore et al. 1997) were found. This incompatibility between models and
observations was pointed out as a shortcoming of current cosmological
models. However, in a recent work by Kravtsov et al. (1997b), who
carefully reanalyzed this problem, they conclude that the internal
structures of DM halos produced in their high resolution N-body
simulations for several CDM models are actually in reasonable
agreement with the observations.

In our approach the dynamical collective processes related to the
substructures are not treated explicitly. Therefore, some of these
processes could be the basis of an effective mechanism of energy
pumping into the central regions of the dark halo. For example, it is
possible that the central part of the halo gains thermal energy due to
the orbital energy loss from dynamical friction of the lumps which are
spiraling in toward the center, and which could be disrupted by tidal
stripping. Within the frame of our approach this effect is compatible
with the introduction of radial perturbations in the orbital motion of
the shells. In a future work this possibility will be studied in
detail. We should note that for a given initial density profile or MAH
a larger tangential velocity component (a larger $e_o$) tends to
reduce the dark halo central density since the orbits of the infalling
particles are more circular. However, if the MAH is such that it
extends continuously to very early epochs as a power law, then, even
with $e_0$=1, the virialized structure will be coreless. On the other
hand, values for $e_o$ larger than $\sim $0.3 (the corresponding 
eccentricity of the orbit is $>0.5$) are not satisfactory as
they imply an extreme situation of dominion of tangential motion with 
respect to the radial motion.

A second vein of exploration in looking for alternatives to produce
cores in the DM structures is related to the cosmological initial
conditions. Having assumed the statistical properties of the
primordial DFF, the most obvious influence of the cosmological model
on the internal structure of the DM halo is related to the nature of
the collisionless particles. If the primordial density fluctuations
are constituted by hot or warm DM particles, then a minimal physical
length scale for the fluctuations will exist. To explore how the
internal structures of virialized objects will change in this case we
use a power spectrum with a cut-off at some minimal scale. It is
reasonable to take this cut-off at the mass scale of the order of
$10^9M_{\odot }$; otherwise, other mechanisms aside from hierarchical
clustering should be invoked in order to form low mass galactic halos.
We assume a toy power spectrum (WDM) which is like the SCDM down to
the scale corresponding to $10^{12}M_{\odot }$ and from this scale
towards smaller masses -if viewed as the mass variance- it flattens
faster than the SCDM in such a way that at $10^9M_{\odot }$ it is
completely flat; scales smaller than $10^9M_{\odot }$ are suppressed.

The DM halos constructed using the WDM-like power spectrum have a
constant density core. Burkert (1995) has pointed out that the
observational data for the dwarf galaxies show some scaling relations
between, for example, the core radius and the rotation velocity at
this radius. To reproduce this kind of correlation and to compare
with the predictions of the WDM-like model, we shall use a fitting
formula for both the observed and theoretically calculated rotation
curves. This formula is based on a density profile of the form:
\begin{equation}
  \rho (r)=\frac{\rho _c}{\left( \frac r{r_c}+a\right) \left( \left(
    \frac r{r_c}\right) ^2+1\right) }
\end{equation}
where $a$ was fixed to a value of $\sim $10. Note that according to
our results the structures of the DM halos are not described by a
single density profile. Nevertheless, as a first approximation, and
bearing in mind that the observational data are still incomplete,
formula (13) is assumed here as universal. The observational data to
be considered consist of the 4 dwarf galaxy rotation curves used in
Moore (1994) and Burkert (1995), the halo rotation curves of the dwarf
galaxy NGC55 derived from the decomposition given in Puche \& Carignan
1991, and the measured rotation curves of the 8 most DM dominated
galaxies of the 17 LSB galaxies presented in de Blok \& McGaugh
(1997). The models were calculated in such a way that a $1-\sigma $
envelope is taken into account. In Figure 5, the maximum rotation
velocity and the core radius obtained from the fitting to the
observational data and the WDM models using formula (13) are plotted. 
The fitting was applied to the increasing part of the rotation curves.
The standard deviation of the fitting formula, scaled to the maximum circular 
velocity given by the same fitting, $\sigma/V_{max}$, is less than $2\%$ 
for all the cases. Consequently, the noise introduced by this fitting 
procedure in the determination of the core size is negligible with 
respect to the observational uncertainties. Dwarf galaxies
show a tight relation between $V_{\max }$ and $r_{core}$, while the
LSB galaxies show a big dispersion in this plane. It is important to
remark that a large observational error is expected in the velocity
measurements of the LSB galaxies, typically associated with beam
smearing. According to Figure 5 when one compares the structures
corresponding to the dwarf galaxies with the DM halo structures
produced in a WDM-like model, this model does not satisfactorily
reproduce the observational trend of $V_{\max}$ with $r_{core}$ in
spite of its ability to generate near constant-density cores. However,
for the LSB galaxies some agreement with the predicted structures is
found, although the scatter is large. A better understanding of the
physics of dwarf galaxies and more observational data on both the LSB
and dwarf galaxies is necessary to reach more definitive conclusions.

\begin{figure}
\plotone{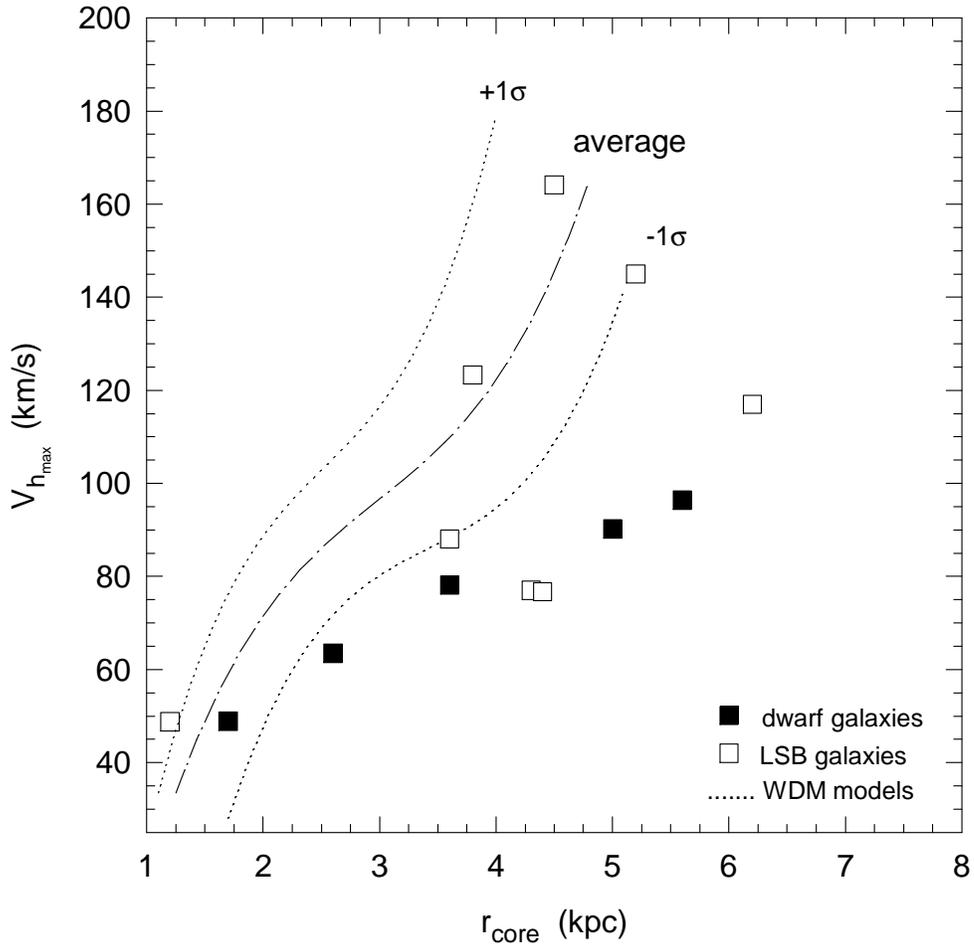}
\caption[fig5.eps]{Core radius vs. maximum halo circular velocity
  for dark halos produced in a WDM like model, and for some dwarf and
  LSB galaxies. In order to estimate the core radii and maximum
  velocities, the observed and calculated rotation curves were fitted
  to a circular velocity profile corresponding to the density profile
  given in eq.(14) with $a\approx $10. The solid line corresponds to
  the average MAHs, while the dotted lines represent the $\pm 1\sigma
  $ deviations. The filled and empty squares correspond to the dwarf
  and LSB galaxies, respectively (see text). \label {fig5}}
\end{figure}

\placefigure{fig5}

The other question related to the ability of the initial cosmological
conditions to influence the internal structure of DM halos is the
statistics of the primordial DFF. The MAHs strongly depend on the
assumed statistical distribution. In the case of Gaussian
fluctuations, for the CDM models the MAHs are such that they
continuously extend to very early epochs giving rise to the steep
inner density profiles. While for the lognormal distribution the
situation may be even worse, other possible statistical distributions
for the primordial DFF might in principle generate MAHs able to
produce non-singular DM halos.

\subsection{The Tully-Fisher relation and its scatter}

It was shown that DM structures emerge from the primordial DFF with
different configurations. Now we shall estimate the statistical
distribution of these configurations, as well as the global relation
between the mass and the gravitational potential of them.

The maximum circular velocity of a DM halo, $V_{h_{\max }}$, is a good
characterization of its gravitational potential. For a given mass,
halos with larger maximum circular velocities are more concentrated,
and most of their matter should have been assembled at earlier times,
when the density of the universe was higher. This directly depends on
the initial conditions, {\it i.e.} the MAHs generated from the
primordial DFF$.$ Given that we know the probability of realization of
every one of the points (mass trajectories) in the plane
$z_1(M_0/2)-z_2(M_0/10)$, we can easily find the probability
distribution of the maximum circular velocities (section 4). In Figure
6 the frequency distributions of maximum circular velocities
corresponding to halos before (a) and after (b) baryon gas collapse
are shown for two masses, $10^{10}M_{\odot }$ and $10^{12}M_{\odot }$,
and for the SCDM model. The velocities were scaled by the average
velocity. It is seen that the distributions are slightly asymmetrical.
The corresponding averages and fractional standard deviations are
given in Table 1. The frequency distributions of maximal circular
velocities are slightly narrower after the dissipative collapse:
initially more concentrated dark halos are less affected by the
gravitational dragging of collapsing dissipative matter than the more
diluted halos.

\begin{figure}
\plotone{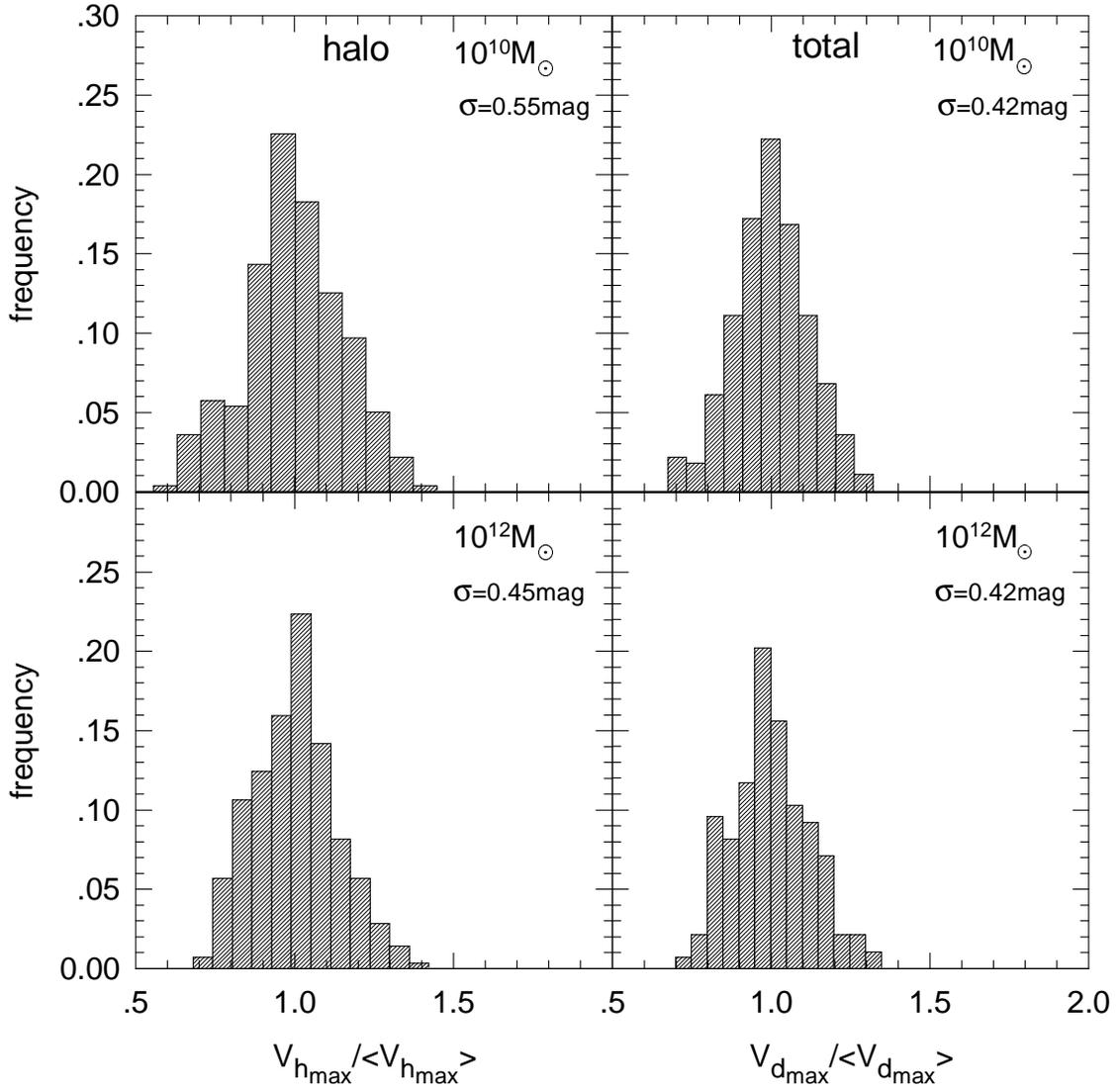}
\caption[fig6.eps]{Frequency distribution of maximum circular
  velocities
(the TF scatter) of halos of $10^{10}M_{\odot }$ (upper panels), and $%
10^{12}M_{\odot }$ (lower panels), before (left panels) and after
(right panels) the gravitational dragging of the baryon matter. The
velocities were normalized to the average velocity (see Table 1). The
fractional velocity deviations, and the scatter in astronomical
magnitudes are given in the boxes. \label{fig6}}
\end{figure}

\placefigure{fig6}

Since for a given mass a range of circular velocities exists, in order
to establish a relationship between $M_0$ and $V_{h_{\max }},$ the
``cosmological'' TF relation, it is necessary to define a typical
circular velocity. Here we shall use the average velocity,
$\overline{V}_{h_{\max }},$ as this typical velocity for a given mass.
This case roughly corresponds to the average MAH. The relation is
$\frac{M_0}{M_{\odot }}=A(\frac{\overline{V}_{h_{\max }}}{km/s})^m ,$
where $A\approx $3.5$\times $10$^4$ , and on average $m \approx $3.38 ($%
m $ very slightly increases with the mass) for the SCDM model, and $%
A\approx $2.3$\times $10$^5$, $m \approx $3.02 for the WDM-like model.
From the analytical estimate of section 2 it was established that the
slope $m $ of the $M_0-V_{h_{\max }}$ relation depends on the slope
$\beta $ of the $M_0-(1+z_f)$ relation. When $\beta \rightarrow $0$,$
then $m \rightarrow $3. The small slope of the ``cosmological'' TF
relation found in our simulations for the SCDM model indeed completely
agrees with the above mentioned small value for $\beta $ ($\sim
$1/15). The relationship between the disk mass $M_d$ and the average
maximum total velocity after the gas collapse, 
$\overline{V}_{t_{\max}}$, $\frac{M_d}{M_{\odot }}=A^{\prime 
}(\frac{\overline{V}_{t_{\max }}}{km/s})^{m ^{\prime }}$, is
nearer to the observable TF relation. After the dissipative collapse
we obtain on the average $A^{\prime }\approx $560, and $m ^{\prime
  }\approx $3.41 for the SCDM model, and $A^{\prime }\approx
$2.0$\times $10$^3$, $m ^{\prime }\approx $3.18 for the WDM-like
model.  The dispersions for both relations can be roughly
characterized by the fractional standard deviations of maximum
circular velocities, $\frac{\sigma _V}{\overline{V}}$ (see Table 1).
They slightly decrease with increasing mass. The fractional standard
deviations in the $M_0-V_{h_{\max }}$ are slightly smaller than those
previously estimated by Eisenstein \& Loeb (1996) who used a much more
simplified scheme to calculate the cosmological TF scatters (note that
they use a SCDM model normalized to $\sigma _8$=1.3).  Using the
slopes of the $M_0-\overline{V}_{h_{\max }}$ and
$M_d-\overline{V}_{t_{\max }}$ relations for the SCDM model the
fractional standard deviations in velocities were translated to
dispersions in masses expressed in astronomical magnitudes (columns 4
and 7 of Table 1). It is interesting to note that after the
dissipative collapse the scatters tend to reduce.

\placetable{tbl-1}

The physical origin of the empirical TF relation is still unclear. The
most general layout of this question deals with the ``nature or
nurture'' dualism, {\it i.e}. was the TF relation established by the
initial conditions of galaxy formation (and, if any, by the same
cosmological conditions), or did it arise as a product of some
evolutionary processes mainly related to the gas hydrodynamics and
star formation history. In the light of the results obtained within
the approach developed here we shall explore whether this relation is
a product or not of the cosmological initial conditions. Since the
galactic luminosity in the redder bands is dominated by older,
low-mass stellar populations, the luminosity in these bands is a
reliable tracer of the disk stellar masses. We shall use the TF
relations in the H band presented in Gavazzi (1993), and in Peletier \&
Willner (1993).  Although the absolute magnitudes for cluster and
field galaxies in Gavazzi (1993) were obtained through aperture
photometry, some corrections were applied in order to obtain an
estimate of the total light emitted by galaxies in the H band (see the
references given in Gavazzi 1993). The line widths were not corrected
for noncircular motions. Therefore, in order to estimate the maximum
rotational velocity, we apply the formula suggested in Tully \&
Fouqu\`{e} (1985), and we normalize it in such a way that the
corrected line widths in the range of 250 to 350 kms$^{-1}$ were in
agreement with the transformations from radio to optical line widths
given by Courteau (1992) and Mathewson et al. (1992). The average
corrections for the different levels at which the line widths are
measured in the different works were taken from Haynes \& Giovanelli
(1984). Peletier \& Willner (1993) presented H band imaging of a
complete sample of galaxies in the Ursa Major cluster and obtained
total magnitudes for these galaxies. Using H I velocity widths from
the compilation of Bottinelli et al. (1990), and applying the
nonrotational correction according to Tully \& Fouqu\`{e} (1985),
Pelletier \& Willner derived the H band TF relation. We normalize
their corrected line widths to the optical observations of Courteau
(1992) and Mathewson et al. (1992), and we assume a distance modulus to
the Ursa Major cluster of 30.95 and a value of H$_0$=85
kms$^{-1}$Mpc$^{-1}$, because this is the value that seems to be in
better agreement with this assumed distance modulus (Pierce \& Tully
1988). Now, in order to transform luminosities into masses a
mass-to-luminosity ratio should be assumed. As Gavazzi et al. (1996)
have shown, in the H band this ratio seems to be constant. They have
calculated from a large sample of galaxies the dynamical
mass-to-luminosity ratio: $\left( \frac M{L_H}\right)
_{dyn}$=2.3$h_{0.5}$ in solar units. Now, using the canonical value of
$\sim $2 for the dynamical-to-visible mass ratio inside the optical
radius in spiral galaxies
(e.g., Rubin 1987), we obtain $\frac{M_d}{L_H}\approx \frac{M_{vis}}{L_H}%
=1.1h_{0.5}$ in solar units. A more direct estimate of this ratio was
made by Thronson \& Greenhouse (1988) using the observed stellar
population in the solar neighborhood. They found a value for the total
stellar mass-to-H luminosity ratio of $\sim $0.55 in solar units, and
they extend this ratio to apply to a wide range of galaxies.
Considering that on average in the disks of spiral galaxies $\sim
$20\% of the mass is contained in gas, the disk mass-to-H luminosity
ratio will be $\sim $0.7. If we assume that the value of the Hubble
parameter is 70 kms$^{-1}$Mpc$^{-1}$, then this mass-to-luminosity
ratio scaling with the Hubble parameter will be $\left(
\frac{M_d}{L_H}\right) _{\odot }=0.5h_{0.5}$. Using M$_{H_{\odot
    }}$=3.46 mag, the TF relations given by Gavazzi (1993) and
Pelletier \&\ Willner (1993) and corrected as explained above, can
be now transformed to the next disk mass-circular velocity relations:
\[
\frac{M_d}{M_{\odot }}=\left\{
\begin{tabular}{c}
  7073 \\ 3215
\end{tabular}
\right\} \frac{V_m^{3.22}}{h_{0.5}}
\]

\[
\frac{M_d}{M_{\odot }}=\left\{
\begin{tabular}{c}
  5216 \\ 2371
\end{tabular}
\right\} \frac{V_m^{3.20}}{h_{0.5}}
\]
where the velocity is given in km/s, and the first and second lines in 
each expression correspond to the Gavazzi et al. (1996) and Thronson \& Greenhouse (1988) mass-to-light
estimates, respectively. These relations are plotted in Figure 7.

For the I band we use the template TF relation presented in Giovanelli
et al. (1997). The mass-to-luminosity ratio in the I band is also not
much dependent on mass or luminosity. If we use the slopes presented
in Gavazzi et al. (1996) for this dependence in the U, B, V and H
bands, and interpolate the corresponding value for the I band, we find
that $\left( \frac M{L_I}\right) _{dyn}\propto M_{dyn}^{0.07}.$
Normalizing this relation in such a way that a typical galaxy
($M_d\approx M_{vis}\approx 5\times 10^{10}M_{\odot })$ has $\left(
\frac M{L_I}\right) _{dyn_{\odot }}\approx $1.8$h_{0.5}$ (Pierce 1990,
taking H$_0$=85 kms$^{-1}$Mpc$^{-1}$), and assuming again a
dynamical-to-visible mass ratio of 2, we obtain $\left(
\frac{M_d}{L_I}\right) _{\odot }\approx 0.9\left( \frac{M_d}{5\times
  10^{10}M_{\odot }}\right) ^{0.07}h_{0.5}$. With this estimate, and
using M$_{I_{\odot }}$=4.19 mag the mass-circular velocity relation
presented in Giovanelli et al. (1997) will be:
\[
\frac{M_d}{M_{\odot }}=\frac{2475}{h_{0.5}^{1.07}}V_m^{3.30}
\]
where the velocity is given in km/s. This relation is also plotted in 
Figure 7. The lower thick solid and
point lines in Figure 7 correspond to predictions for SCDM $\sigma
_8=1,$ and WDM-like models, respectively. Note that the mass is scaled
with the Hubble parameter. Strictly speaking, the curves corresponding
to the models should not be plotted in this graphic since they were
calculated only for one value of the Hubble parameter ($h_{0.5}$=1).
Nevertheless, it is easy to show that the model curves in Figure 7
will have roughly the same behavior with $h_{0.5}$ as the
observational curves. Since $V\propto r_v^{-1/2},$ and according to
the top-hat model $r_v\propto h_{0.5}^{\beta -2/3},$ where $\beta $ is
related to the power spectrum index, then $V\propto h_{0.5}^{1/3-\beta
  /2}.$ Using for the SCDM model $\beta \approx $1/15 (see above),
then $V\propto h_{0.5}^{0.3}.$ In the case of the WDM model $\beta $
is even smaller than this value. For the observational curves the
velocity roughly scales with the Hubble parameter as $h_{0.5}^{1/m}$
where $m\approx $3.2-3.3. Therefore, in Figure 7 the observational and
theoretical curves scale with $h_{0.5}$ roughly in the same way.

\begin{figure}
\plotone{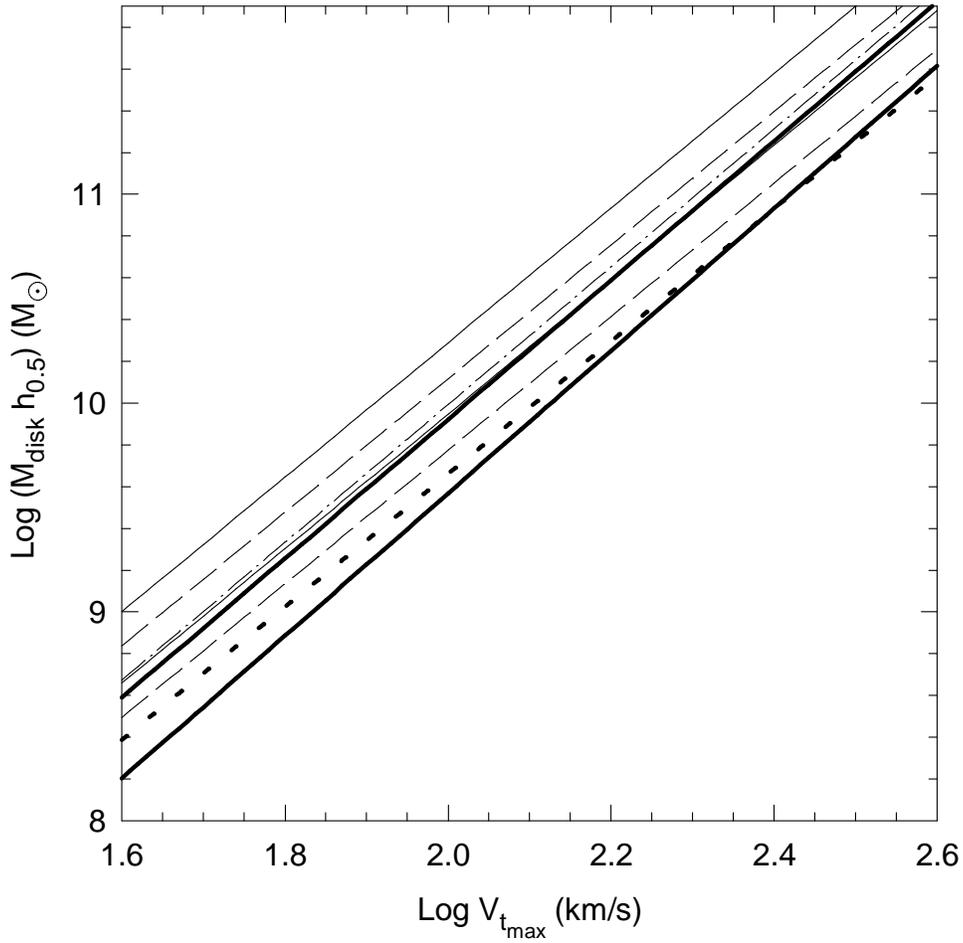}
\caption[fig7.eps]{The thin solid and dashed lines represent the
  H$-$band TF relation from Gavazzi (1993) and from Peletier \& Willner
  (1993), respectively.  Both split by two after this relation was
  transformed to the $M_d-V_{d_{\max }}$ relation using the two
  mass-to luminosities ratios given in the text.  The maximum
  velocities were inferred from the HI line widths presented by these
  authors and corrected for the effects of line broadening due to gas
  random motions (see text). The dot-dashed line corresponds to the
  I$-$band
TF relation presented by Giovanelli et al. (1997), and transformed to the $%
M_d-V_{d_{\max }}$ relation. The thick lower solid line is for the
$\sigma _8=1$ SCDM model, while the thick upper solid line is for the
same model but with the power spectrum normalized to $\sigma _8=0.57$.
The thick short-dashed line is the prediction for the WDM like model.
\label{fig7}}
\end{figure}

\placefigure{fig7}

The mass-velocity relations theoretically obtained for the SCDM and
WDM-like models have slopes very similar to the ones inferred from
observations.  However, as it is seen in Figure 7, these models,
specially the CDM case, produce objects with too high circular
velocities. This is not a surprise since it is well known that the
SCDM $\sigma _8=1$ model suffers from an excess of power at small
scales. We have calculated models for the same SCDM case, but taking
$\sigma_8=0.57,$ a value suggested by the masses and abundances of
rich clusters of galaxies (White, Efstathiou, \& Frenk 1993).  The
obtained mass-velocity relation is plotted in Figure 7 (upper thick
solid line). The curve is clearly better centered in the range of the
observational estimates, and it is very near to the curves we consider
most reliable. An extrapolation to galactic scales of the
observational estimates for the galaxy distribution power spectrum
(see fig. 4 in Stompor et al. 1996, and references therein) is also in
rough agreement with a SCDM power spectrum normalized to $\sigma
_8\approx 0.5-0.7$. We argue that at the basis of this consistency is
the fact that {\it the TF\ relation is a manifestation at galactic
  scales of the same mass distribution observed at larger scales, and
  primarily established by the initial cosmological conditions, i.e.
  the power spectrum of fluctuations. }It is worth noting that the
$\sigma _8\approx 0.6$ SCDM power spectrum at galactic scales roughly
resembles the {\it COBE} normalized power spectrum of a flat $\Lambda
$CDM model with $\Omega _\Lambda =$0.7 and $h=$0.7.

The question of the scatter in the TF relation has been widely
discussed in the literature (for a review see Strauss \& Willick
1995). Recent studies by Willick et al. (1997) (see for the other
references therein), and by Mathewson \& Ford (1994) using large and
complete samples, found scatters in the I band of 0.35-0.40 mag and
0.42 mag, respectively; in the H band a scatter of 0.45 mag was found
(see the reference in Strauss \& Willick 1995). The scatter estimated
in Giovanelli et al. (1997) for the I band varies between $\sim $0.25
mag among fast rotators to $\sim $0.40 mag among the slowest rotators
of their sample. Eisenstein \& Loeb (1996) have shown that for open or
flat, low-$\Omega _0$ CDM models the scatter is lower than for the
SCDM model. The ratio of the scatters of the flat, $\Omega _\Lambda
=$0.7, and the SCDM models which they calculated is roughly 0.7.
Hence, the scatter of $\sim $0.42 mag predicted here for the SCDM
model could be reduced to $\sim $0.3 mag for the flat, $\Omega
_\Lambda =$0.7 model.

To estimate the influence of different $\lambda ^{\prime }s$ on the
results obtained here, we have calculated models with $\lambda =$0.1
for the SCDM case. In the range of $10^{10}-10^{12}M_{\odot }$ the
slope of the TF relation remains the same as for the $\lambda =$0.05
case, while the zero-point increases by a factor of $\sim $1.73 which
corresponds to $\sim $0.57 mag. Taking this value as a standard
deviation -which can be considered as an upper limit since the
intrinsic distribution of $\lambda $ is commonly approximated by a
lognormal distribution with $\langle {\lambda }\rangle\approx $0.05
and $\sigma _\lambda <1$ (e.g., Catelan \& Theuns 1996)- and
considering that the dispersions in $\lambda $ and in the MAHs are
independent -again this will imply an upper limit estimate- the
standard deviation of the SCDM TF relation would result be $\sigma
_{tot}=\sqrt{\sigma ^2(MAH)+\sigma ^2(\lambda )}\approx $0.71 mag,
while for the $\Lambda $CDM model it could be $\approx $0.64 mag. If
these theoretical estimates of the TF scatter are intended to be
compared with the observational data, then the last should consider
also the family of LSB galaxies since the effect of changing $\lambda
$ directly reflects on the surface brightness and scale lengths of
disk galaxies.

\section{Conclusions}


1. In a hierarchical clustering scenario the shapes of DM virialized
structures depend on mass and on the history of mass aggregation of
the systems, ranking from very compact configurations to nearly
isothermal structures. The range of MAHs for an isolated galaxy of a
given mass follows a distribution determined by the statistical
properties of the primordial DFF. The average MAHs corresponding to a
given mass produce DM halos with a density profile well described by
the outcomes of N-body cosmological simulations (NFW96,$\ $NFW97,
Kravtsov et al. 1997a). Since our approach is based on the assumption
of spherical symmetry, adiabatic invariance, and only a moderate
influence of the external tidal forces on the gravitational collapse,
such an agreement reveals the robustness of these assumptions, at
least in what concerns the external structure of DM halos.

2. The inner density profiles of dark halos produced in the SCDM model
(and probably in other CDM models) are slightly
shallower than $\rho \propto r^{-1}.$ Direct (rotation curves of
dwarf and LSB galaxies) and indirect (rotation curve decompositions of
normal galaxies) observational data do not confirm this result.
Collective dynamical processes related to the collapsing substructures
which are considered only in a parametric and limited fashion in our
approach, may be at the heart of this inconsistency. These processes
might be able to generate efficient mechanisms of energy pumping to
the central regions of the dark halos.  Comparisons with the results
of high resolution N-body simulations will show whether these
processes are important or not. Unfortunately the results of the
N-body simulations are still controversial. Introducing a cut-off at
low wavelengths in the power spectrum of fluctuations, {\it i.e}.
using a WDM-like model, we have found that the virialized structures
have nearly constant density-profile cores, although the size of these
cores tend to a constant value as the maximum circular velocities of
the system increase.

3. Within the frame of the heuristic approach given in section 2 the
slope of the TF relation is easily connected to the slope of the
nominal mass-formation redshift relation. In the more realistic
context of hierarchical clustering, galaxies are continuously forming
according to their MAHs which determine the structures of their DM
halos. Since the MAHs depend on mass and on the statistical properties
of the primordial density fluctuations, a structural relation between
total mass and the maximum circular velocity of the system (the
cosmological TF relation), with a natural scatter, arises. For a SCDM
model we predict $\frac{M_0}{M_{\odot }}=$3.5$\times $10$^4(\frac{\overline{V}_{h_{\max }}}{km/s})^{3.38}$ with a fractional 
standard deviation in velocity,
$\frac{\sigma _V}{\overline{V}_{h_{\max }}}$, of $\sim $16\% and $\sim
$13\% for $10^{10}M_{\odot }$ and $10^{12}M_{\odot }$ systems,
respectively. For the WDM-like model used here, we obtained
$\frac{M_0}{M_{\odot }}=$2.3$\times $10$^5(\frac{\overline{V}_{h_{\max }}}{km/s})^{3.02}$ and the deviations are smaller than for the SCDM model.

4. The dissipative collapse of the baryon gas within the virialized DM
halos increases the concentration of the halos in the central regions. The
factor by which the maximum circular velocity increases after the
dissipative collapse, taking $f$=$f_B$=0.05 and $\lambda $=0.05, for
the SCDM model, oscillates roughly between 1.30 and 1.43. The
dissipative collapse tends to reduce the initial scatter in the
mass-maximum circular velocity relation: for both the $10^{10}M_{\odot
  }$ and $10^{12}M_{\odot }$ SCDM galaxies the resulting standard
deviations in velocity are $\sim $12\%, or 0.42 mag.  The disk
mass-maximum rotation velocity relation for the SCDM model is
$\frac{M_d}{M_{\odot }}=$560$\times (\frac{\overline{V}_{t_{\max }}}{km/s})^{3.41}$, while that for the WDM-like model is 
$\frac{M_d}{M_{\odot }}=$2.0$\times $10$^3(\frac{\overline{V}_{t_{\max}}}{km/s})^{3.18}$. The slopes of these relations are in good agreement with the observational estimates.

5. Although the maximum circular velocities of the structures after
baryon collapse decrease with increasing the spin parameter $\lambda
$, the slope of the TF relation remains the same. The theoretical TF
scatter (an upper limit) estimated as the independent combination of
the dispersions in the MAHs and in $\lambda $ increases to $\sim $0.7
mag for the SCDM model.

6. The $\sigma _8=$1 SCDM model produces galaxies with high rotation
velocities as compared to observations (Figure 7). Using a SCDM power
spectrum normalized to $\sigma _8=0.57$, the theoretical prediction is
in excellent agreement with the most reliable observational estimates
of the H and I band TF relations. The observational estimates of the space galaxy distribution are also in agreement with the SCDM power spectrum 
with this normalization (at galactic scales this power spectrum is similar to 
that of the flat, {\it COBE} normalized, $\Lambda $CDM model with $\Omega_\Lambda=$0.7 and $h=$0.7). In this way, the H and I band TF 
relations can be considered as the natural extension to galactic scales of 
the galaxy distribution power spectrum, which reveals that the TF relation is mainly a product of the power spectrum of fluctuations and the 
rate at which the protogalaxies aggregate mass, {\it i.e.} at the basis of 
its nature are the cosmological initial conditions. The H and I band TF relations can be used to probe the power spectrum at galactic scales.

In a forthcoming paper the galactic evolution of the disks built up in
the evolving dark halos will be calculated with the aim to explore the
possibility that the morphological Hubble sequence is a direct result
of the cosmological initial conditions, and to find the parameters on
which this sequence depends.

\acknowledgments

V.A.-R. received financial support through the project UNAM-DGAPA
IN-105894. He also gratefully acknowledges a fellowship from the
program ``Becas Cuauhct\'{e}moc'' of CONACyT. C.F and V.A.-R.  would 
like to acknowledge to Antonio Garc\'{e}s for his computing assistence.

\begin{center}

  References
\end{center}

\begin{description}
\item Albada, T.S. van, Bahcall, J.N., Begeman, K., \& Sancisi, R.
  1985, \apj, 295, 305

\item Avila-Reese, V. 1998, PhD. Thesis, UNAM

\item Avila-Reese, V., \& Firmani, C. 1997, in ``Dark and visible
  matter in galaxies and cosmological implications'',{\it \ }eds.
  Persic, M., \& Salucci, P., ASP Conference Series, vol. 117, 416

\item Bardeen, J.M., Bond, J.R., Kaiser, N., \&\ Szalay, A.S. 1986,
  \apj, 304, 15

\item Bernstein, G.M., Guhathakurta, P., Raychaudhury, S., Giovanelli,
  R., Haynes, M.P., Herter, T. \& Vogt, N.P. 1994, \aj, 107, 1962

\item Bond, J.R., Cole, S., Efstathiou, G., \& Kaiser, N. 1991, \apj,
  304, 15

\item Bond, J.R., \& Myers, S.T. 1996, \apjs, 103, 1

\item Bottinelli, L., Gouguenheim, L., Fouqu\`{e}, P., \& Paturel, G.
  1990, A\&AS, 82, 391

\item Bower, R.G. 1991, \mnras, 248, 332

\item Burkert, A. 1995, \apj, 447, L25

\item Burstein, D., \& Sarazin, C.L. 1983, \apj, 264, 427

\item Carlberg, R., \& Couchman, H.M.P. 1989, \apj, 219, 18

\item Catelan, P., \& Theuns, T. 1996, \mnras, 282, 436

\item Cole, S.M. \& Lacey, C. 1996, \mnras, 281, 716

\item Crone, M., Evrard, A.E., \& Richstone, D.O. 1994, \apj, 434, 402

\item Courteau, S. 1992, Ph.D. Thesis, University of California, Santa
  Cruz

\item Dalcanton, J.J., Spergel, D.N., \& Summers, F.J. 1996, \apj, 482, 659
\item de Blok, W.J.G. 1997, in ``Dark and visible matter in galaxies
  and cosmological implications''{\it , }eds. Persic, M., \& Salucci,
  P., ASP Conference Series, vol. 117, 39

\item de Blok, W.J.G., \& McGaugh, S.S. 1997, \apj, \mnras, 290, 533

\item de Vaucouleurs, G., \& Olson, D.W. 1982, \apj, 256, 346

\item Dubinsky, J., \& Carlberg, R.G. 1991, \apj, 378, 496

\item Eisenstein, D.J., \& Loeb, A.1996, \apj, 459, 432

\item Faber, S.M., \& Jackson, E.J. 1976, \apj, 204, 668

\item Fall, S.M., \& Efstathiou, G. 1980, \mnras, 193, 189

\item Flores, R.A., \& Primack, J.R. 1994, \apj, 427, L1

\item Flores, R.A., Primack, J.R., Blumenthal, G.R., \& Faber, S.M.
  1993, \apj, 412, 443

\item Fukushige, T., \& Makino, J. 1996, \apj, 477, L9

\item Gavazzi, G. 1993, \apj, 419, 469

\item Gavazzi, G., Pierini, D., and Boselli, A. 1996, \aap, 312, 397

\item Giovanelli, R., Haynes, M.P., Herter, T.H., Vogt, N.P., da
  Costa, L.N., Freudling, W., Salzer, J.J., \& Wegner, G. 1997, \aj,
  113, 53

\item Gott, J.R. 1975, \apj, 201, 296

\item Gott, J.R., \& Rees, M.J. 1975, \aap, 45, 365

\item Gunn, J.E. 1977, \apj, 218, 592

\item Gunn, J.E., \& Gott, J.R. 1972, \apj, 176, 1

\item Haynes, M.P., \& Giovanelli, R. 1984, \aj, 89, 758

\item Hoffman, Y. 1988, \apj, 328, 489

\item Hoffman, Y., \& Shaham, J. 1985, \apj, 297, 16

\item Jedamzik, K. 1995, \apj, 448, 1

\item Katz, N. 1991, \apj, 368, 325

\item Kravtsov, A.V., Khoklhov, A.M. \& Klypin, A.A. 1997a, ApJS, 111,
  73

\item Kravtsov, A.V., Klypin, A.A., Bullock, J., \& Primack, J. 1997b,
  preprint (astro-ph/9708176)

\item Kuhlman, B., Melott, A.L., \& Shandarin, S.F. 1996, \apj, 470,
  L41

\item Lacey, C., \& Cole, S. 1993, \mnras, 262, 627

\item Manrique, A. \& Salvador-Sol\'{e}, E. 1995, \apj, 453, 6

\item \underbar{\hskip 1.2truecm}. 1996, \apj, 467, 504

\item Mathewson, D.S., Ford, V.L. 1994, \apj, 434, L39

\item Mathewson, D.S., Ford, V.L., \& Buchhorn, M. 1992, \apjs, 81,
  413.

\item Melott, A.L., Shandarin, S.F., Splinter, R.J., \& Suto, Y. 1997,
  \apj , 479, L79

\item Moore, B. 1994, \nat, 370, 629

\item Moore, B., Governato, F., Quinn, T., Stadel, J., \& Lake, G.
  1997, preprint, (astro-ph/9709051).

\item Navarro, J., Frenk, C.S., \& White, S.D.M. 1996, \apj, 462, 563
  (NFW96)

\item \underbar{\hskip 1.2truecm}. 1997, \apj, 490, 493 (NFW97)

\item Peacock, J.A., \& Heavens, A.F. 1990, \mnras, 243, 133

\item Peebles, P.J.E. 1969, \apj, 155, 393

\item \underbar{\hskip 1.2truecm}. 1980, ``The Large Scale Structure
  in the Universe'' (Princeton University Press, Princeton)

\item \underbar{\hskip 1.2truecm}. 1984, \apj, 277, 470

\item Peletier, R.F., \& Willner, S.P. 1993, \apj, 418, 626

\item Pierce, M.J. 1990, in ``Evolution of the Universe of Galaxies'',
  ed.  Kron, R.G. (ASP vol.10, Provo, Utah), p.48

\item Pierce, M.J., \& Tully, R .B. 1988, \apj, 330, 579

\item Press, W.H., \& Schechter, P. 1974, \apj, 187, 425

\item Puche, D., \& Carignan, C. 1991, \apj, 378, 487

\item Rees, M.J., \& Ostriker, J.P. 1977, \mnras, 179, 541

\item Rubin, V.C. 1987, in IAU Symp. 117, ``Dark Matter in the
  Universe'', eds. J. Kormendy \& G. Knapp (Dordrecht:\ Reidel), p.51

\item Ryden, B.S. 1988, \apj, 329, 589

\item Ryden, B.S. \& Gunn, J.E. 1987, \apj, 318, 15

\item Silk, J. 1977, \apj, 211, 638

\item \underbar{\hskip 1.2truecm}. 1984, \nat, 301, 574

\item Stompor, R., Gorski, K.M., \& Banday, A.J. 1996, \mnras, 277,
  1225

\item Strauss, M.A., \& Willick, J.A. 1995, \physrep, 261, 271

\item Thuan, T.X., \& Gott, J.R. 1977, \apj, 216, 194

\item T\'{o}th, G., \& Ostriker, J.P. 1992, \apj, 389, 5

\item Thronson, H. A., \& Greenhouse, M.A. 1988, \apj, 327, 671

\item Tully, R.B., \& Fisher, J.R. 1977, \aap, 54, 661

\item Tully, R.B., \& Fouqu\`{e}, P. 1985, \apjs, 58, 67

\item Voglis, N., Hiotelis, N., \& Harsoula M. 1995, \apss, 226, 213

\item Warren, M.S., Quinn, P.J., Salmon, S.K., \& Zurek, W.H. 1992,
  \apj, 399, 405

\item Willick, J.A., Courteau, S., Faber, S.M., Burstein, D., Dekel,
  A., \& Strauss, M.A. 1997, \apjs, 109, 333

\item White, S.D.M. 1984, \apj, 286, 38

\item \underbar{\hskip 1.2truecm}. 1994, preprint MPA 831

\item White, S.D.M., \& Rees, M.J. 1978, \mnras, 183, 341

\item White, S.D.M., Efstathiou, G., \& Frenk, C.S. 1993, \mnras, 262,
  1023

\item Zaroubi, S., \& Hoffman, Y. 1993, \apj, 416, 410

\item Zaroubi, S., Naim, A., \& Hoffman, Y. 1996, \apj, 457, 50

\end{description}

\begin{deluxetable}{crrrrrr}
  \footnotesize \tablecaption{Average maximum velocities and
    dispersions for the $SCDM$ model.
\label{tbl-1}}
\tablewidth{0pt} \tablehead{ \colhead{Mass\tablenotemark{a}} &
  \colhead{$\overline{V}_{h_{\max}}$ \tablenotemark{b}} &
  \colhead{$\frac{\sigma_V}{\overline{V}_{h_{\max}}}$} &
  \colhead{$\sigma_h(mag)$} & \colhead{$\overline{V}_{t_{\max}}$
    \tablenotemark{c}} &
  \colhead{$\frac{\sigma_V}{\overline{V}_{t_{\max }}}$} &
  \colhead{$\sigma_t(mag)$} } \startdata 1 &41.0 &0.16 &0.55 &55.7
&0.12 &0.42\nl 100 &161.4 &0.13 &0.45 &215.0 &0.12 &0.42\nl
 
\enddata

\tablenotetext{a}{in units of $10^{10}M_{\odot }$}
\tablenotetext{b}{before the baryonic collapse, in $Kms^{-1}$}
\tablenotetext{c}{after the baryonic collapse, in $Kms^{-1}$}

\end{deluxetable}

\clearpage

\end{document}